\documentclass[aps,prl,twocolumn,superscriptaddress,floatfix,nofootinbib]{revtex4-1}

\usepackage{graphicx}
\usepackage{amssymb}
\usepackage{amsmath}
\usepackage[dvipsnames]{xcolor}
\begin{document}

\title{Entropic Timescales of Dynamic Heterogeneity in Supercooled Liquid}
\author{Vinay Vaibhav}
\email{vinay.vaibhav@unimi.it}
\affiliation{The Institute of Mathematical Sciences, CIT Campus, Taramani, Chennai 600113, India}
\affiliation{Department of Physics “A. Pontremoli”, University of Milan, Via Celoria 16, 20133 Milan, Italy}
\author{Suman Dutta}
\email{suman.dutta@icts.res.in (Corresponding Author)}
\affiliation{International Centre for Theoretical Sciences - Tata Institute of Fundamental Research\\
Survey No 51, Hessaraghatta Hobli, Shivakote, Bangalore 560089, India}

\begin{abstract}
Non-Gaussian displacement distributions are universal predictors of dynamic heterogeneity in slowly varying environments. Here, we explore heterogeneous dynamics in supercooled liquid using molecular dynamics simulations and show the efficiency of the information-theoretic measure in quantifying dynamic heterogeneity over the widely used moment-based quantifications of non-Gaussianity. Our analysis shows that the heterogeneity quantified by the negentropy is significantly different from the one obtained using the conventional approach that considers deviation from Gaussianity up to lower-order moments. Further, we extract the timescales of dynamic heterogeneity using the two methods and show that the differential changes diverge as the system experiences strong intermittency near the glass transition.  
\end{abstract}                   
\maketitle

{\it Introduction.---} Fickian theory of diffusion has been unquestionably successful for more than a century in analyzing particle-level dynamics in soft condensed matter that appear in different forms and shapes unless the system is intermittent or has widely separating timescales \cite{glotzer1997, edigerRev, blaaderen2000,weitz, gao,richertRev, eric0, glotzer2003, andersen2005, shell2005, szamel2010, garrahan2011, kirsten, berthier2011, berthier_book2011, berthierBiroliRev2011, manning2013, sastryChandanRev, szamel2014, ploymerHetro2017, karmakar2018, karmakar2020, saroj2023}. In metastable systems, the fundamental route to diffusion becomes difficult in the presence of complex energy landscapes, specifically, when a small magnitude of thermal fluctuation is not enough to supply the energy cost of achieving diffusion, overcoming the energy barriers \cite{berthier_book2011,heuer}. For instance, molecular displacements deviate from their usual Gaussian form \cite{blaaderen2000, gao} in liquids approaching glass transition \cite{pinaki2007, pinaki2008}, showing slow heterogeneous density relaxation \cite{ngcage}. Such dramatic slowing down observed generically in a host of systems without any reproducible thermodynamic transition has remained a surprise even after decades of research \cite{edigerRev, berthierBiroliRev2011, sastryChandanRev}. Dynamic heterogeneity is the observed complex dynamics of particles in such temporally fluctuating environments in the presence of spatial degrees of heterogeneity, where both locally fast and slow relaxation processes coexist simultaneously \cite{chandan2010, berthier_book2011, eric, karmakar2018, dutta2019}. 

Dynamic heterogeneity in the supercooled liquid has been affirmed with persisting non-Gaussian tails in the displacement distributions, even when the mean-squared displacement linearly increases with time \cite{granick2009,granick2012,berthier2022}. Such class of non-Gaussian diffusion has been explained as an effective dynamics in the presence of a diffusion spectrum where the dynamics is strongly influenced by the presence of {\it cages} and diffusion is only restored upon {\it cage-breaking}, making it distinct from the Fickian class of liquids \cite{granick2009,granick2012}. The persistence of dynamic heterogeneity is, thus, debated for the presence of multiple timescales of relaxation and various ways of their determination. It has also remained indecisive whether the onset of diffusion at all occurs in the finite time when a liquid approaches its glass transition \cite{berthier2011, burov2020, berthier2022,pastore}.

Extracting the fundamental timescale of dynamic heterogeneity directly from the displacement distributions or the {\it self-Van Hove} function is advantageous than obtaining it by other quantities which are either related to its moments or derivatives. Using the displacement distributions, it is possible to identify timescales of heterogeneity by finding the maximal non-Gaussianity using conventional measures that rely on moment-based relationships \cite{kobBook}. However, one hindrance of this method is that such moment ratios are primarily limited to lower-order moments, which raises natural questions: {\it How optimal are the moment-based predictions of dynamic heterogeneity or its detection by the conventional techniques?} This calls for newer directions in the precise identification of non-Gaussianity by informative approaches with the data-driven resources, like the information-theory-based optimizations simplify challenging multi-scale and inverse problems in bio-informatics \cite{shell2016,barbier} or the prediction of dynamics and structures using machine intelligence is unveiling newer avenues in physics \cite{andrea2015,deepmind2020,berthier2023}.

\begin{figure*}[t]
  \includegraphics[height=8.0cm]{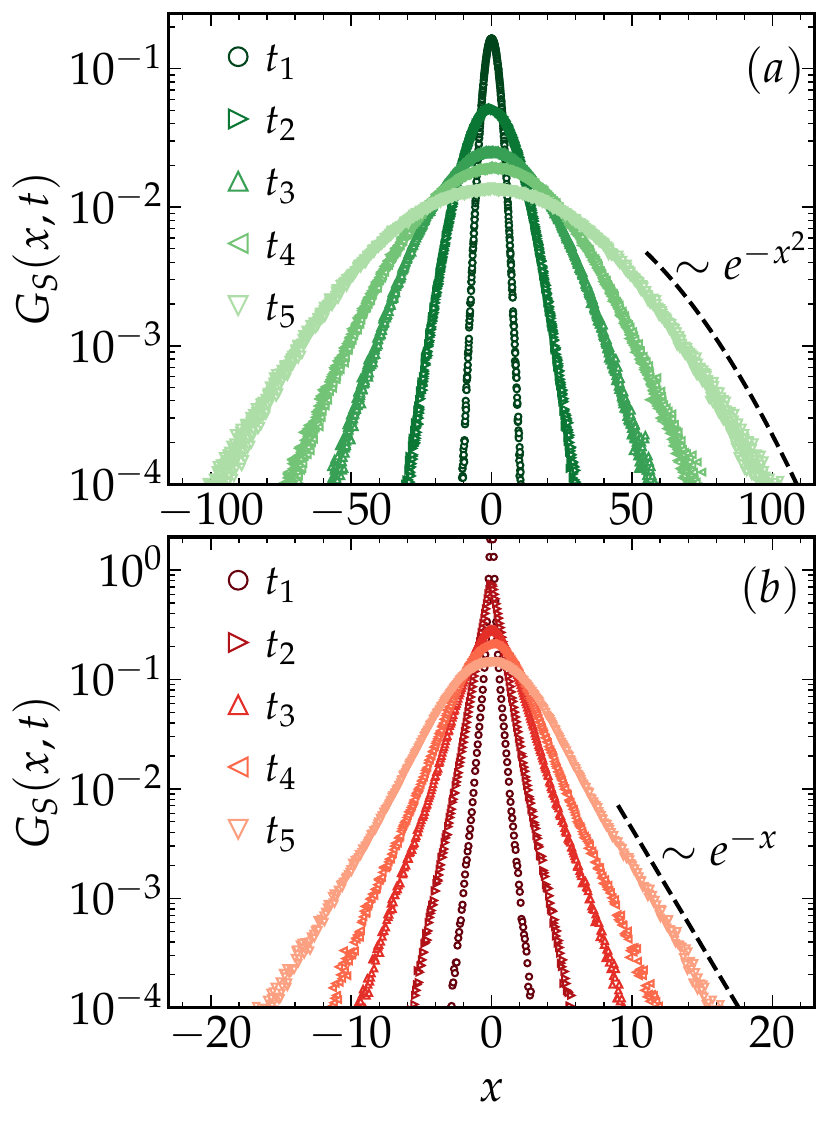}
  \includegraphics[height=8.0cm]{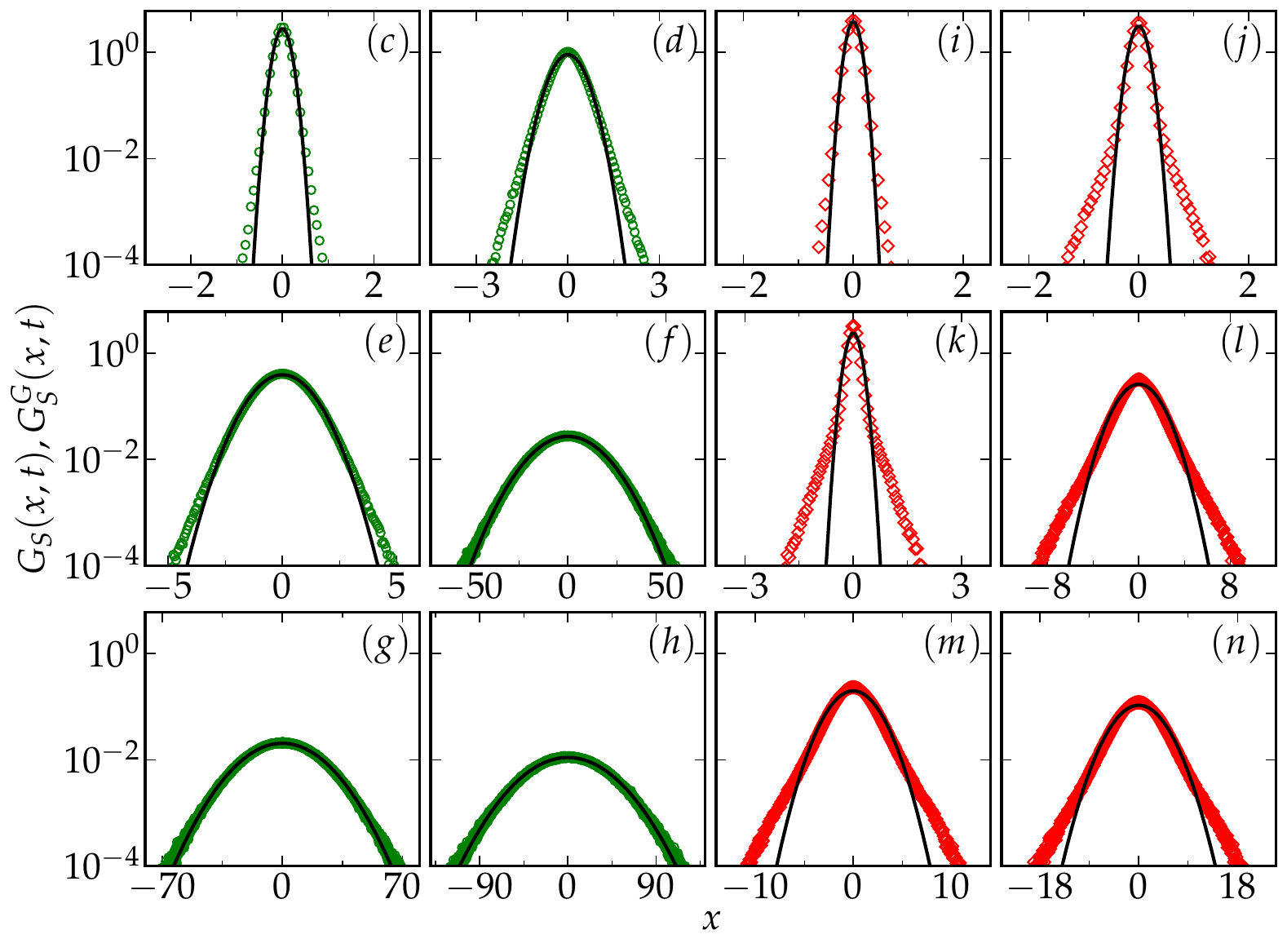}
  \caption{[(a)-(b)] Temporal evolution of the Non Gaussian displacement distributions,  $G_{S}(x,t)$ vs. $x$ for two different $T$, (a) $T=0.70$ (b) $T=0.45$ for different $t$: $t_{1}=928.98$, $t_{2}=9527.62$, $t_{3}=39913.85$, $t_{4}=68300.79$, $t_{5}=139796.16$ with $x\in [X,Y,Z]$. Long time behavior of $G_{S}(x,t)$ at large $x$ is Gaussian for $T=0.70$ and exponential for $T=0.45$. [(c)-(n)] Temporal evolution of the displacement distributions, $G_{S}(x,t)$ (open symbols) and respective reconstructed equal-time nearest Gaussians, $G^{G}_{S}(x,t)$ (black solid line) vs. $x$ for $T=0.70$ [(c)-(h)] and $T=0.45$ [(i)-(n)] for $t=$ $0.60$ [(c),(i)], $21.62$ [(d),(j)], $129.59$ [(e),(k)], $27899.00$ [(f),(l)], $47740.94$ [(g),(m)] and $167210.14$ [(h),(n)]. $G^{G}_{S}(x,t)$ are the optimal Gaussians with identical first two moments same as $G_{S}(x,t)$.}
  \label{fig1}
\end{figure*}

 Here, we explore dynamic heterogeneity in the supercooled liquid above the glass transition using molecular dynamics simulations in three dimensions. We examine the spatio-temporal dynamics in terms of the evolving probability distribution functions of the displacements at single particle-level which are strongly heterogenous and non-Gaussian till it diffuses at sufficiently long time. We quantify temporal intermittency in terms of non-Gaussianity extracted from the molecular displacements using the conventional moment-based descriptions and a {\it relative entropy} based {\it non-Gaussian information} that considers the statistical distance between the time-dependent displacement distribution and its equal-time nearest Gaussian trained from the original probability distribution functions. We extract and compare the identified timescales of optimal heterogeneity obtained from the two methods and show that they surprisingly differ in estimating microscopic heterogeneity, in particular when the situation is strongly intermittent at low temperatures. Further, we show that such difference diverge when approaching the glass transition while the timescales are similar at relatively high temperatures within the supercooled regime. We correlate the two quantities and interpret the deviation. 

{\it Model.---} We simulate a well studied model glass-forming system, popularly known as the Kob-Anderson 80:20 (A:B) binary mixture \cite{kob1995testing}, in three dimensions at different temperatures $T$ within the supercooled regime. The particles (each with unit mass) interact via the Lennard-Jones (LJ) pair-potential with a cutoff at a distance $R_{c}$. Here, we use $N = 1000$ particles  at a constant number density ($\rho = 1.20$) in a cubic box of length $L=9.41$ and $R_{c} = 2.5$. All measurements are done in the LJ reduced unit (see Ref. \cite{vaibhav2020} for model related information). Independent molecular dynamics trajectories are generated using LAMMPS \cite{lammps} under periodic boundary conditions with integration time step $\Delta t = 0.004$ at each of the different $T\in [0.45, 0.70]$, above the mode-coupling temperature $T_{MCT} \approx 0.435$ of the mixture \cite{kob1995testing}.

{\it Results.---} We track the particle trajectories and compute time-dependent probability distribution functions of particle displacements (also known as self-van Hove functions) which have been mapped in one dimension, $x$ in time interval, $t$, $G_s(x,t) =\frac{1}{N} \sum_{i=1}^N \delta(x-(x_i(t)-x_i(0)))$ \cite{vHove} by averaging over displacement distributions in all spatial dimensions obtained in different independent trajectories. 

We show the time evolution of $G_{S}(x,t)$ for different $t$ in Fig.~\ref{fig1} for two different values of $T$. At finite temperatures, $G_{S}(x,t)$ spreads with increasing $t$ due to diffusion which we see here. For $T=0.70$, it develops non-Gaussian tail for intermediate times that reverts back to Gaussian at long times, suggesting that the dynamics is intermittent and heterogeneous even at relatively high temperatures, within the supercooled regime [Fig.~\ref{fig1}(a)]. For $T=0.45$, the heterogeneity is not only stronger, but also its lifetime is prolonged as the non-Gaussian tail ($G_{S}(x,t)\sim \exp(-x/\lambda(t))$ with $\lambda(t)\sim \sqrt{t}$) persists for very long time [Fig.~\ref{fig1}(b)]. Such persistent exponential tails have been earlier reported in both simulations \cite{shell2005,pinaki2007} and experiments \cite{blaaderen2000,berthier_book2011,granick2012}. 

In order to assess the degree of temporal heterogeneity in microscopic dynamics, we show the temporal evolution of $G_{S}(x,t)$ for various $t$ for two different $T$ in Figs. ~\ref{fig1}(c)-\ref{fig1}(n) and compare it with the reconstructed {\it equal-time nearest Gaussian distributions} (see Ref. \cite{sumanEPL2020}), $G^{G}_{S}(x,t)$ that has first two moments same as $G_{S}(x,t)$. This ensures that for a purely Gaussian distribution,  $G^{G}_{S}(x,t)\rightarrow G_{S}(x,t)$ and $G^{G}_{S}(x,t)\ne G_{S}(x,t)$ when it is strictly non-Gaussian. We observe $G_{S}(x,t)\approx \delta (x)$ for $t=0$, but the calculation of $G^{G}_{S}(x,t)$ will become more meaningful when $G_{S}(x,t)$ has some finite support at finite $t$. For $T=0.70$, we observe that for very small $t$, the difference between $G_{S}(x,t)$ and $G^{G}_{S}(x,t)$ is not very much significant [Fig.~\ref{fig1}(c)]. The difference enhances with increasing $t$ [Figs.~\ref{fig1}(d), \ref{fig1}(e)] where $G_{S}(x,t)$ is prominently non-Gaussian while for larger $t$, again $G_{S}(x,t)$ reverts back to Gaussian form with $G^{G}_{S}(x,t)\approx G_{S}(x,t)$ [Figs.~\ref{fig1}(f)-\ref{fig1}(h)], suggesting the onset of Fickian diffusion that is characterized by the presence of Gaussian displacement distribution and linear mean squared displacement \cite{granick2012}.

\begin{figure}[t]
\includegraphics[width=8.0 cm]{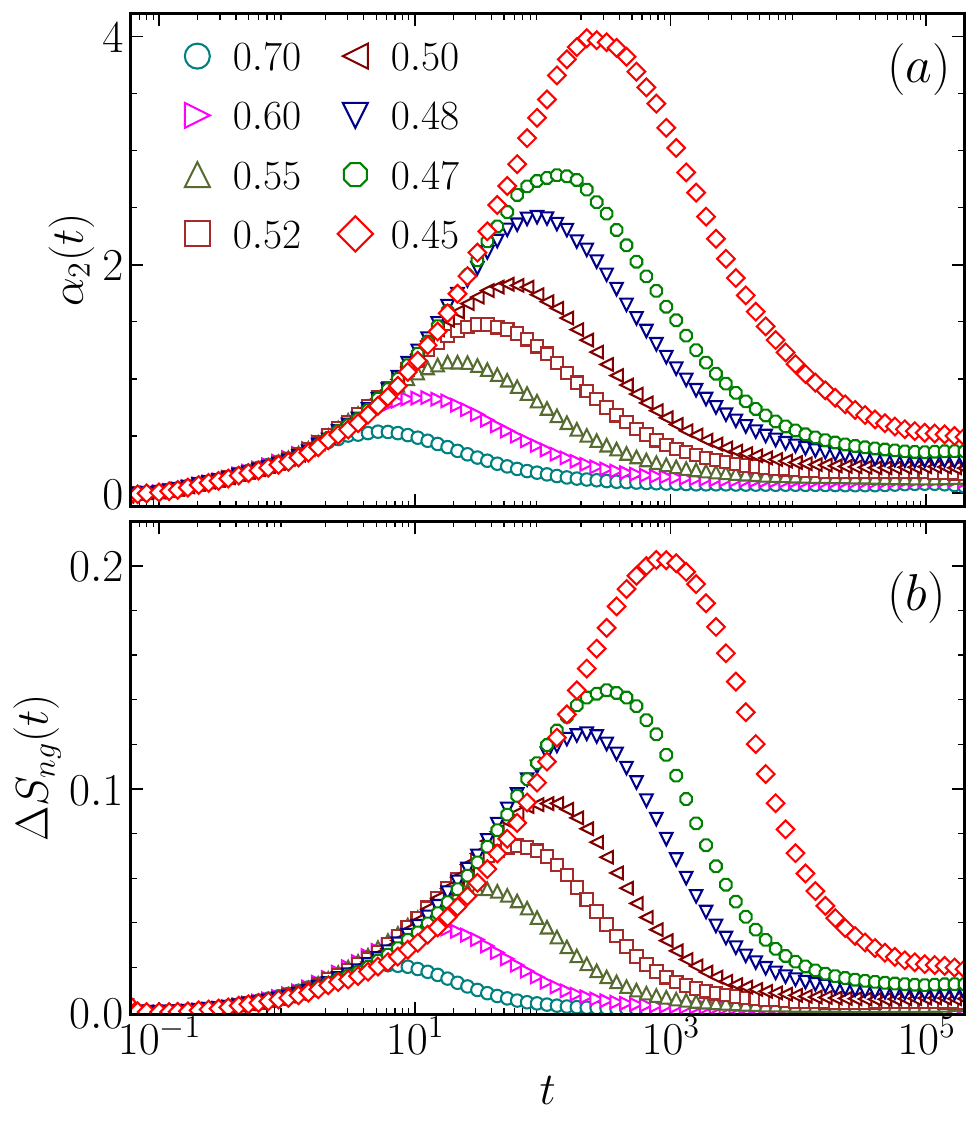}
\caption{Quantification of dynamic heterogeneity in terms of non-Gaussianity of $G_{S}(x,t)$ via (a) $\alpha_{2}$ and (b) $\Delta S_{ng}$ vs. $t$ for different $T$ as marked.}
\label{fig2}
\end{figure}

The situation is surprisingly different for $T=0.45$ which we show in Figs.~\ref{fig1}(i)-\ref{fig1}(n) and compare it with respective $G^{G}_{S}(x,t)$, as earlier. For small $t$, the distributions are highly spiked, centered at $x=0$. Within the smaller intervals, $G_{S}(x,t)\approx G^{G}_{S}(x,t)$ [Fig.~\ref{fig1}(i)] and smaller deviations from delta functions are mostly contributed by the lower-order moments. For larger $t$, as $G_{S}(x,t)$ broadens in $x$, more deviation could be seen from $G^{G}_{S}(x,t)$ in the form of exponential tail that grows with a significantly large proportion of displacements appearing beyond the support of the equal time nearest Gaussians [Figs.~\ref{fig1}(j), \ref{fig1}(k)], suggesting that the heterogeneity is maximally contributed by the higher-order moments. With increasing $t$ [Figs.~\ref{fig1}(l)-\ref{fig1}(n)], the difference between $G_{S}(x,t)$ and $G^{G}_{S}(x,t)$ decreases yet it remains non-Gaussian within our observation time-window.

Now we attempt to quantify the dynamic heterogeneity in terms of non-Gaussianity in $G_{S}(x,t)$ using conventional approaches that use lower order moment-based relationships. One such very simple quantification of non-Gaussianity that is widely available in the literature is based on the deviation of kurtosis from the square of second moment, defined as $\alpha_2 (t) =\frac{\langle x^{4}(t) \rangle}{3\langle x^{2}(t) \rangle^2} - 1$ with $<x^{n}(t)>=\int dx \ x^{n} G_{S}(x,t)$, as in Ref. \cite{kob1995testing, glotzer1997, horbach1998, blaaderen2000}. We show the dependence of $\alpha_{2}$ with $t$ for different $T$ in Fig.~\ref{fig2}(a). For $T=0.70$, $\alpha_{2}$ grows with increasing $t$ till a peak is observed. After that, $\alpha_{2}$ decreases for larger $t$. For smaller temperatures, the peak in $\alpha_{2}$ shifts to larger $t$ and the height of the peak grows. In all these cases, $\alpha_{2}$ decreases monotonically after the peak. The timescale corresponding to the peak is generally identified as the characteristic timescale of dynamic heterogeneity. However, such quantification is fundamentally limited to fourth-order moments of $G_{S}(x,t)$. Figs.~\ref{fig1}(c)-\ref{fig1}(n) suggest that the higher order moments may be associated with the larger degrees of heterogeneity for which we explore another quantification of non-Gaussianity that captures contributions of all order moments of $G_{S}(x,t)$ at a given time.

Non-Gaussianity in $G_{S}(x,t)$ is now analysed using {\it non-Gaussian information} \cite{sumanEPL2020} that we developed following Ref. \cite{simon2012}, originally proposed as {\it Negentropy}. It uses the statistical distance between $G_{S}(x,t)$ and $G^{G}_{S}(x,t)$ to quantify $\Delta s_{ng}(t) = A~ D_{KL} (G_{S} (x,t)|| G^{G}_{S} (x,t) )$ where $A$ is a constant. We further define, $\Delta S_{ng}(t) (={\Delta s_{ng}(t)}/{A}) = -\int dx ~ G_{S} (x,t) \log_{e} \frac{G^{G}_{S}(x,t)}{G_{S}(x,t)} = S^{G}_{ng}- S_{ng}$ when $G_{S}(x,t)$ has some finite support. Here, $S_{ng}(t)=-\int~dx~G_{S}(x,t)\log_{e}G_{S}(x,t)$ and  $S^{G}_{ng}(t)=-\int~dx~G^{G}_{S}(x,t)\log_{e}G^{G}_{S}(x,t)$ and $D_{KL}(P||Q)$ is the {\it Kullback-Leibler (KL)} divergence \cite{kl1951} between two probability distribution functions, $P(x)$ and $Q(x)$. We show the time dependence of $\Delta S_{ng}$ for different $T$ in Fig.~\ref{fig2}(b). $\Delta S_{ng}$ shows non-monotonic dependence on $t$, similar to $\alpha_{2}$. For all these cases, $\Delta S_{ng}$ grows up to a peak and then decreases monotonically. Surprisingly, for every $T$, we observe that $\Delta S_{ng}$ still grows to higher values when $\alpha_{2}$ has already reached the maximum. This suggests that the dynamic heterogeneity in $G_{S}(x,t)$ may not be completely captured in $\alpha_{2}$ and the peak in $\alpha_{2}$ may also not be a true representation of timescale of the underlying heterogeneity as it underestimates the contribution of the higher order moments in $G_{S}(x,t)$ because the information in $\alpha_{2}$ is limited up-to the fourth moment of the displacement distribution.

\begin{figure}[t]
\includegraphics[width=8.5cm]{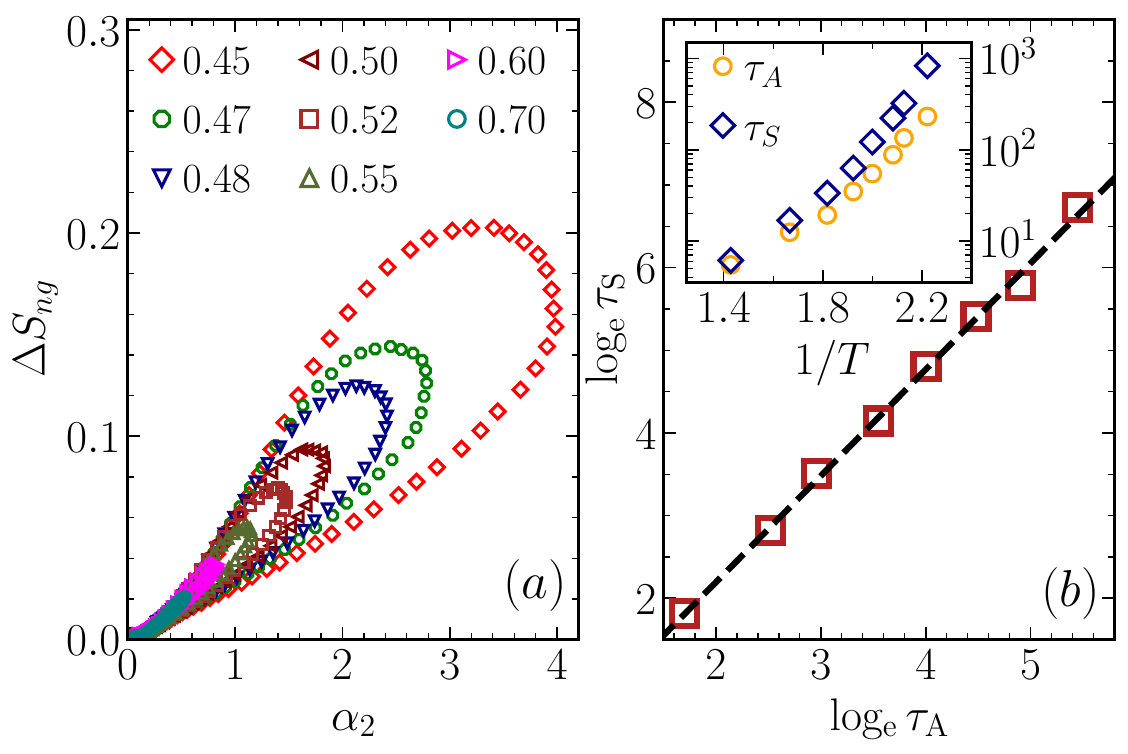}
\includegraphics[width=8.5cm]{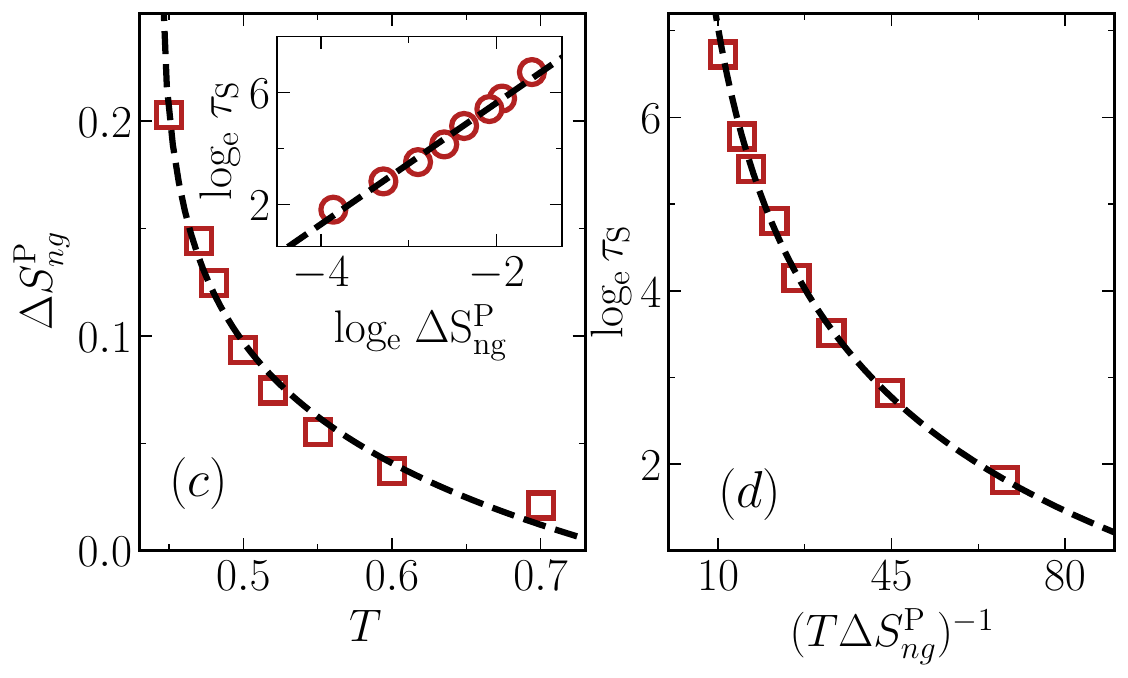}
\caption{(a) Correlating the non-Gaussian information, $\Delta S_{ng}$ with the non-Gaussian parameter, $\alpha_{2}$ for different $T\in [0.45,0.7]$. (b) Inset: Temperature dependence of the timescales extracted corresponding to the peak of $\alpha_{2}$, $\tau_{A}$ and $\Delta {S}_{ng}$, $\tau_{S}$. Main Panel: $\log_{e} \tau_{S}$ vs. $\log_{e} \tau_{A}$. The dashed line shows a linear fit: $ \log_{e}\tau_{S} \sim \beta_{0} + \beta \log_{e}\tau_{A}$. (c) Dependence of $\Delta S^{P}_{ng}$ on $T$. The dashed line shows $\Delta S^{P}_{ng} \sim \nu_0 - (T-T*)^{\gamma}$. Inset: $\log_{e}\tau_{S}$ vs. $\log_{e} \Delta S^{P}_{ng}$. The dashed line shows $\log_{e}\tau_{S}\sim \zeta_{0}+ \zeta \log_{e} \Delta S^{P}_{ng}$. (d) $\log_{e} \tau_{S}$ vs. $(T \Delta S^{P}_{ng})^{-1}$. The dashed line shows $\log_{e}\tau_{S}\sim (T\Delta S^{P}_{ng})^{-\psi}$. See text for the values of different fitting parameters.}
\label{fig3}
\end{figure}

We compare the behavioural differences between $\alpha_{2}$ and $\Delta S_{ng}$ in Fig.~\ref{fig3}. In Fig.~\ref{fig3}(a), we statistically correlate $\alpha_{2}$ and $\Delta S_{ng}$ for the different cases of $T$. In all these cases, they form loop-like shapes and the area covering the curve is larger for smaller $T$. We observe that the behavioral difference between $\alpha_{2}$ and $\Delta S_{ng}$ grows when both have higher values, representing the maximal heterogeneity. The degree of such deviation significantly increases for decreasing $T$, suggesting strongly that the dynamic heterogeneity is underestimated by $\alpha_{2}$ when $G_{S}(x,t)$ is maximally non-Gaussian. This is consistent with our earlier investigations \cite{sumanEPL2020} where we obtained similar loops in the case of a model supercooled liquid based on continuous time random walk (CTRW), suggesting that the underlying dynamic heterogeneity picture is qualitatively similar and the behavior of non-Gaussianity based quantifications remain universal under such slowly varying conditions.

We further extract the associated timescales corresponding to the peaks of $\alpha_{2}$, $\tau_{A}$ and that of $\Delta S_{ng}$, $\tau_{S}$ [see Fig.~\ref{fig2}] and show its dependence on inverse of $T$ in the inset of Fig.~\ref{fig3}(b). Both $\tau_{A}$ and $\tau_{S}$ increase sharply for larger values of $1/T$. In all these cases, we observe that $\tau_{S}$ has higher values than that of $\tau_{A}$. We finally correlate $\log_{e} \tau_{A}$ and  $\log_{e} \tau_{S}$ in the main panel of Fig.~\ref{fig3}(b). We fit the data of $\log_{e} \tau_{S}$ and $\log_{e} \tau_{A}$ and obtain $\log_{e} \tau_{S} \approx \beta_{0} + \beta \log_{e} \tau_{A}$ with $\beta_{0} \approx -0.379, \beta \approx 1.286$. This suggests that $\tau_{S}$ diverges from $\tau_{A}$ in a power-law ($\tau_{S} \sim \tau_{A}^{\beta}$) and such difference diverges with decreasing $T$. Further, we extract the height of the peak in $\Delta S_{ng}$, $\Delta S^{P}_{ng}(T) (=\max_{t\in [0,\infty]} \Delta S_{ng}(t;T))$ and correlate it with the corresponding entropic timescale, $\tau_{S}$. In inset of Fig. 3(c), we show that $\log_{e}\tau_{S}\sim \zeta_{0}+ \zeta \log_{e} \Delta S^{P}_{ng}$ where $\zeta_{0}\approx 9.956, \zeta\approx 2.164$. Also, we observe that $\Delta S^{P}_{ng}$ increases with decreasing $T$, sharply near $T = 0.45$ [Fig.~\ref{fig3}(c)], when $\tau_{S}$ grows sharply as shown in inset of Fig.~\ref{fig3}(b). We fit the data as $\Delta S^{P}_{ng} \sim \nu_{0} - (T-T*)^{\gamma}$ with $T*\approx 0.443, \nu_{0} \approx 0.928, \gamma\approx 0.064$. This suggests that the {\it non-Gaussian information} diverges as $T\rightarrow T*$ for the sharp increase of $\tau_{S}$ when approaching to the state of dynamic arrest at the glass transition. This also indicates that the degree of heterogeneity at a given $T$ is strongly connected with $\Delta S^{P}_{ng}$. Thus, the system explores a strongly intermittent environment and dynamically heterogeneous states with larger non-Gaussian information when the divergence of $\Delta S^{P}_{ng}$ represents the absolute entropic distance of the dynamic arrest from the nearest diffusive route. Hence, the term $T\Delta S^{P}_{ng}$ can be considered as an energy term associated with the heterogeneity that competes with the diffusion in overcoming the dynamic heterogeneity at a given $T$. So, we correlate the entropic timescale, $\tau_{S}$ with $1/(T\Delta S^{P}_{ng})$ and observe that $\tau_{S}$ grows and shows a sharp rise with decreasing $1/(T\Delta S^{P}_{ng})$ [Fig.~\ref{fig3}(d)]. Our data follows $\log_{e}\tau_{S}\sim \eta_{0} + \eta(T\Delta S^{P}_{ng})^{-\psi}$ with $\eta_{0}\approx-8.993, \eta \approx 25.691, \psi \approx -0.205$. Using the form of $\Delta S^{P}_{ng}$, one can obtain $\log_{e} \tau_{S} \sim T^{-\psi}[1-(T-T*)^{\gamma}]^{-\psi}$, where the divergence of $\tau_{S}$, for $T\rightarrow T*$, is primarily governed by the term $[1-(T-T*)^{\gamma}]^{-\psi}$, which qualitatively explains the nature of growth of $\tau_{S}$ with decreasing $T$, as seen in inset of Fig. 3(b). The rapid rise of $\tau_{S}$ is due to the presence of strong intermittency, for which achieving diffusion becomes increasingly difficult overcoming the dynamic heterogeneity, for the exploration within the rough energy landscapes at low temperatures.

{\it Discussion.---} These results align with our theoretical predictions of the entropic timescales of dynamic heterogeneity in a model supercooled liquid \cite{sumanEPL2020} where the development of intermittent non-Gaussian tail was modeled using the {\it Montroll-Weiss} CTRW framework, considering complex hopping of particles within the {\it mobile} and {\it immobile} regions and jumps from one region to another \cite{langerMukh2008}. The persistence of such vibrations and jumps within these regions overall controls the nature of intermittency and its lifetime \cite{pinaki2007, pinaki2008}. Such intermittent non-Gaussian tails were inferred as a convolution process of cooperative diffusion, known as the {\it Brownian yet Non-Gaussian diffusion} \cite{granick2012, sokolov}. Supportive literature \cite{slater, metzler, jain2016, dutta2016, karmakar2018, karmakar2020} shows that the diffusion spectrum obtained upon the deconvolution of the non-Gaussian displacement distribution validates the physical picture of dynamic heterogeneity which considers the simultaneous presence of {\it slow} and {\it fast} regions within the system \cite{dutta2019}. Such heterogeneity exhibits anomalous spatiotemporal fluctuations that have all order information \cite{Lemaitre}. Therefore, to estimate the dynamic heterogeneity effectively, consideration of all order moments is advantageous. Gaussianity reverts when the Fickian diffusion sets in and the dynamics follow the central limit theorem, marking the onset of diffusion. However, extraction of this timescale is difficult due to the scale-free decay of $\alpha_{2}(t)$ \cite{pastore}, specifically, when the ambient temperature is close to the Glass transition and the situation becomes intrinsically non-equilibrium. Our analysis also affirms that both $\tau_{A}$ and $\tau_{S}$ diverge while the latter diverges faster than the former in a power law which can be further tested using other CTRW models \cite{pinaki2007,pinaki2008} or using Mode-coupling-theory \cite{ho}. The sharp increase of $\Delta S^{P}_{ng}$ with decreasing $T$ is due to diverging dynamic heterogeneity that also leads to a sharp increase in entropic timescale as the system approaches the state of dynamic arrest. Whether the non-Gaussian-information has any connection with the configurational entropy \cite{sastryjack} or non-equilibrium free energy\cite{information00}, needs further investigations.  

{\it Conclusions.---} We quantify the timescales of dynamic heterogeneity in supercooled liquids using the conventional {\it non-Gaussian parameter} and the {\it non-Gaussian information}. We show that the entropic timescales are significantly different and diverging from those obtained using the non-Gaussian parameter. This difference arises because the moment-based definitions are limited up to the fourth order, while the information-theoretic quantification takes into account all order moments. Although several other moment-based definitions are available \cite{szamel2006, horbach1998}, it is always challenging to estimate timescales using them as the computation of higher-order moments can be increasingly noisy or unreliable without the quality data. On the other hand, our framework is easy to implement and extracts heterogeneity optimally. This makes the information-theoretic-framework scientifically robust in quantifying non-Gaussianity in practical situations, also in a more general context, in out-of-equilibrium systems in identifying or predicting novel cross-over or transition where small fluctuations lead to catastrophic changes, or differentiate phases or states of matter \cite{Lemaitre, loewen}.

{\it Acknowledgements.---} V.V. gratefully acknowledges funding from the European Union through Horizon Europe ERC Grant number: 101043968 “Multimech”. S.D. acknowledges support of the Department of Atomic Energy, Government of India, under project no RTI4001. We acknowledge HPC facilities at IMSc, ICTS-TIFR and supporting grants,  ICTS/eiosm2018/08, ICTS/ISPCM2023/02. We thank D. Bagchi and S. Sikdar for sharing data. We thank S. Bose, P. Chaudhuri, C. Dasgupta, J. Horbach, S. Karmakar, S. K. Nandi, M. S. Shell and A. Zaccone for insightful discussions. 

{\it Research Contribution.---} VV: Performed simulations and formal analysis, Validation, Visualizations. SD: Conceptualization, Methodology, Validation, Research Coordination, Data Interpretation, Writing. 
  


\begin{thebibliography}{99}

\bibitem{glotzer1997}
W. Kob, C. Donati, S. J. Plimpton, P. H. Poole, and S. C. Glotzer,
Phys. Rev. Lett. {\bf 79}, 2827 (1997).
%

\bibitem{edigerRev}
M. D. Ediger,
Annu. Rev. Phys. Chem. {\bf 51}, 99 (2000).
%

\bibitem{blaaderen2000}
W. K. Kegel and A. van Blaaderen, 
Science {\bf 287}, 290 (2000).
%

\bibitem{weitz}
E. R. Weeks, J. C. Crocker, A. C. Levitt, A. Schofield, and D. A. Weitz, 
Science {\bf 287}, 5453 (2000).
%

%
\bibitem{gao}
Y. Gao and M. L. Kilfoil,
Phys. Rev. Lett. {\bf 99}, 078301 (2008).
%

\bibitem{richertRev}
R. Richert,
J. Phys.: Condens. Matter {\bf14}, R703 (2002).
%

\bibitem{eric0}
E. Bertin, J.-P. Bouchaud, and F. Lequeux,
Phys. Rev. Lett. {\bf 95}, 015702 (2005).
%

\bibitem{glotzer2003}
N. Lačević, F. W. Starr, T. B. Schrøder, and S. C. Glotzer,
J. Chem. Phys. {\bf 119}, 7372 (2003).
%

\bibitem{andersen2005}
H. C. Andersen,
Proc. Natl. Acad. Sci. {\bf 102}, 6686 (2005).
%

\bibitem{shell2005}
M. S. Shell, P. G. Debenedetti, and F. H. Stillinger,
J. Phys.: Condens. Matter {\bf 17} S4035 (2005).
%

\bibitem{szamel2010}
E. Flenner and G. Szamel,
Phys. Rev. Lett. {\bf 105}, 217801 (2010).
%

\bibitem{garrahan2011}
J. P. Garrahan, 
Proc. Natl. Acad. Sci. {\bf 108}, 4701 (2011).
%

\bibitem{kirsten}
K. Martens, L. Bocquet, and J-L. Barrat, 
Phys. Rev. Lett. {\bf 106}, 156001 (2011).
%

\bibitem{berthier2011}
L. Berthier, 
Physics {\bf 4}, 42 (2011).
%

\bibitem{berthier_book2011}
L. Berthier, G. Biroli, J.-P. Bouchaud, L. Cipeletti, and W. van Saarloos, 
Oxford University Press, Oxford, (2011).
%

\bibitem{berthierBiroliRev2011}
L. Berthier and G. Biroli,
Reviews of modern physics, {\bf 83}(2), 587 (2011).
%

\bibitem{manning2013}
E. M. Schoetz, M. Lanio, J. A. Talbot, and M. L. Manning,
Journal of The Royal Society Interface {\bf 10}, 20130726 (2013).
%

\bibitem{sastryChandanRev}
S. Karmakar, C. Dasgupta, and S. Sastry,
Annu. Rev. Condens. Matter Phys. {\bf 5}(1), 255 (2014).
%

\bibitem{szamel2014}
E. Flenner and G. Szamel,
Phys. Rev. Lett. {\bf 112}, 097801 (2014).
%

\bibitem{ploymerHetro2017}
W. Zhang, J. F. Douglas, and F. W. Starr,
J. Chem. Phys {\bf 146}, 203310 (2017).
%

\bibitem{karmakar2018}
B. P. Bhowmik, I. Tah, and S. Karmakar,
Phys. Rev. E {\bf 98}, 022122 (2018).
%

\bibitem{karmakar2020}
R. Das, C. Dasgupta, and S. Karmakar,  
Frontiers in Physics {\bf 8}, 210 (2020).
%

\bibitem{saroj2023}
P. Pareek, M. Adhikari, C. Dasgupta, and S. K. Nandi,
arXiv preprint arXiv:2305.18042 (2023).
%

%
\bibitem{heuer}
A. Heuer, 
J. Phys.: Condens. Matter {\bf 20}, 373101 (2008).
%

\bibitem{pinaki2007}
P. Chaudhuri, L. Berthier, and W. Kob,
Phys. Rev. Lett. {\bf 99}, 060604 (2007).
%

\bibitem{pinaki2008}
P. Chaudhuri, Y. Gao, L. Berthier, M. Kilfoil, and W. Kob,
J. Phys.: Condens. Matter {\bf 20}, 244126 (2008).
%

%
\bibitem{ngcage}
B. Vorselaars, A. V. Lyulin, K. Karatasos, and M. A. J. Michels, 
Phys. Rev. E {\bf 75}, 011504 (2007).
%

%
\bibitem{chandan2010}
S. Karmakar, C. Dasgupta, and S. Sastry,
Phys. Rev. Lett. {\bf 105}, 015701 (2010).
%

%
\bibitem{eric}
G. Gradenigo, E. Bertin, and G. Biroli, 
Phys. Rev. E {\bf 93}, 060105(R) (2016).
%

\bibitem{dutta2019}
S. Dutta,
Chem. Phys. {\bf 522}, 256 (2019).
%

\bibitem{granick2009}
B. Wang, S. M. Anthony, S. C. Bae, and S. Granick,
Proc. Natl. Acad. Sci. {\bf 106}(36), 15160 (2009).
%

\bibitem{granick2012}
B. Wang, J. Kuo, S. C. Bae, and S. Granick,
Nat. Mater. {\bf 11}, 481 (2012).
%

\bibitem{berthier2022}
L. Berthier, E. Flenner, and G. Szamel, 
arXiv preprint arXiv:2210.07119 (2022).
%

%
%

\bibitem{burov2020}
E. Barkai and S. Burov,
Phys. Rev. Lett. {\bf 124}(6), 060603 (2020).
%

\bibitem{pastore}
F. Rusciano, R. Pastore, and F. Greco, 
Phys. Rev. Lett. {\bf 128}, 168001 (2022).
%

%
\bibitem{kobBook}  
K. Binder and W. Kob,
Glassy materials and disordered solids: An introduction to their statistical mechanics. World Scientific, 2011.
%

%
\bibitem{shell2016}
M. S. Shell, 
Coarse-Graining with the Relative Entropy, edited by Rice S. A. Rice and A. R. Dinner (John
Wiley) 2016.
%

%
\bibitem{barbier}
M. Gabrie, A. Manoel, C. Luneau, J. Barbier, N. Macris, F.
Krzakala, and L. Zdeborova, 
Advances in Neural Information Processing Systems {\bf 31}, 1826 (2018).
%

\bibitem{andrea2015}
E. D. Cubuk, S. S. Schoenholz, J. M. Rieser, B. D. Malone, J. Rottler, D. J. Durian, E. Kaxiras, and A. J. Liu,  
Phys. Rev. Lett. {\bf 114}, 108001 (2015).
%

\bibitem{deepmind2020}
V. Bapst, T. Keck, A. Grabska-Barwińska, C. Donner, E. D. Cubuk, S. S. Schoenholz, A. Obika, A. W. Nelson, T. Back, D. Hassabis, and P. Kohli,
Nat. Phys. {\bf 16}, 448 (2020).
%

\bibitem{berthier2023}
G. Jung, G. Biroli, and L. Berthier, 
Phys. Rev. Lett. {\bf 130}, 238202 (2023).
%

\bibitem{kob1995testing}
W. Kob and H. C. Andersen, 
Phys. Rev. E {\bf 51}, 4626 (1995).
%

\bibitem{vaibhav2020}
V. Vaibhav, J. Horbach, and P. Chaudhuri,
Phys. Rev. E {\bf 101}(2), 022605 (2020).
%

%
\bibitem{lammps} 
S. Plimpton,  
J. Comp. Phys. {\bf 117}, 1 (1995).
%

\bibitem{vHove}
L. Van Hove,
Phys. Rev. {\bf 95} 249, (1954).
%

\bibitem{sumanEPL2020}
R. Dandekar, S. Bose, and S. Dutta,
Europhys. Lett. {\bf 131}, 18002 (2020).
%

\bibitem{horbach1998}
J. Horbach, W. Kob, and K. Binder, 
Philos. Mag. B {\bf 77}, 297 (1998).
%

\bibitem{simon2012}
J. S. Ivan, M. S. Kumar, and R. Simon, 
Quantum Inf. Process. {\bf 11} 853 (2012).
%

\bibitem{kl1951}  
S. Kullback and R. A. Leibler,
Ann. Math. Stat. {\bf 22}, 79 (1951).
%

\bibitem{langerMukh2008}
J. S. Langer and S. Mukhopadhyay, 
Phys. Rev. E {\bf 77}, 061505 (2008).
%

\bibitem{sokolov}
A. V. Chechkin, F. Seno, R. Metzler, and I. M. Sokolov, 
Phys. Rev. X {\bf 7}, 021002 (2017).
%

\bibitem{dutta2016}
S. Dutta and J. Chakrabarti, 
Europhys. Lett. {\bf 116}, 38001 (2016).
%

\bibitem{slater}
M. V. Chubynsky and G. W. Slater, 
Phys. Rev. Lett. {\bf 113}, 098302 (2014).
%

\bibitem{metzler}
V. Sposini, A. Chechkin, and R. Metzler, 
J. Phys. A: Math. Theor. {\bf 52}, 04LT01 (2018).
%

\bibitem{jain2016}
R. Jain and K. L. Sebastian,
J. Phys. Chem. B {\bf 120}, 9215 (2016).
%

\bibitem{Lemaitre}
E. Lemaitre, I. M Sokolov, R. Metzler, and A. V Chechkin, 
New. J. Phys. {\bf 25}, 013010 (2023).
%

\bibitem{ho}
S.-H. Chong, 
Phys. Rev. E {\bf 78}, 041501 (2008).
%

\bibitem{sastryjack}
F. W. Starr, J. F. Douglas, and S. Sastry, 
J. Chem. Phys. {\bf 138}, 12A541 (2013).
%

\bibitem{information00}
J. M. R. Parrondo, J. M. Horowitz and T. Sagawa, Nat. Phys., {\bf 11}, 131 (2015)

\bibitem{szamel2006}
G. Szamel and E. Flenner, 
Phys. Rev. E {\bf 73}, 011504 (2006).
%

\bibitem{loewen}
X. Zheng, B. ten Hagen, A. Kaiser, M. Wu, H. Cui, Z. Silber-Li, and H. Löwen,
Phys. Rev. E {\bf 88}, 032304 (2013).
%

\end{thebibliography}
\end{document}